\documentclass[12pt]{article}
\usepackage{graphicx}
\usepackage{epsfig}
\usepackage{graphics}
\usepackage{booktabs}
\usepackage{cite}
\usepackage{array}
\usepackage{longtable}
\usepackage{pdflscape}
\usepackage{hyperref}
\usepackage{subfigure}

 \usepackage{fullpage}

\begin{document}

\author{Arvind W. Kiwelekar, Sanil S. Gandhi, Laxaman D. Netak,\\  Shankar B. Deosarkar \thanks{Department of Computer Engineering, Dr. Babasaheb Ambedkar Technological University, Lonere-402103, Raigad(M.S.), India.}}

 \title{Use-cases of Blockchain Technology for Humanitarian Engineering}
 
This is a pre-print of the following chapter: Arvind W. Kiwelekar, Sanil S Gandhi, Laxman D. Netak, Shankar B Deosarkar, Use cases of blockchain technology for humanitarian engineering, published in  Information and Communication Technologies for Humanitarian Services,2020,
edited by Muhammad Nazrul Islam, 2020, publisher IET. The authenticated version is available on \href{https://digital-library.theiet.org/content/books/10.1049/pbte089e_ch7}{publisher site}

{\bf Cite this chapter as:}
Arvind W Kiwelekar, Sanil Gandhi, Laxman D Netak and Shankar B Deosarkar  Use-cases of Blockchain Technology for Humanitarian Engineering in   Information and Communication Technologies for Humanitarian Services(pp143-163)  Publisher: IET Publication. Editor Muhammad Nazrul Islam,

\newpage
\maketitle
\section*{Abstract}Humanitarian Engineers need innovative methods to make
technological interventions for solving societal problems. 
The emerging blockchain technology has the enormous potential to provide effective interventions in various developmental sectors, including Agriculture, Education, Health, and Transportation. 
In these sectors, mediators have been considered as one of the impediments for developmental work. 
Blockchain technology facilitates peer-to-peer business transactions, thus eliminating the role of mediators.
Hence, the blockchain technology is emerging as an alternative to conventional mediator-centred solutions adopting client-server based Internet technologies.

A combination of blockchain technology with other technologies can be used to address domain-specific challenges. 
For example, the combination of blockchain technology and Internet-of-Thing (IoT) has the potential to monitor the usage of scarce resources such as the level of ground-water and amount of energy consumption. 

The aims of this chapter are twofold. Firstly, it describes the primary building blocks of blockchain technology. Secondly, it illustrates various use-case scenarios of blockchain technology in the fields of Agriculture, Energy Health and others.
 
\section{Humanitarian Engineering: An Example}

The goal of Humanitarian Engineering is to help underprivileged and marginalized people.  
It does so by designing and implementing technology-based solutions to address the challenges faced by such people. The {\em Jaipur foot} \cite{arya2008jaipur} is one of the best examples of humanitarian engineering products. The Jaipur foot is an artificial leg designed to help people with below-knee leg amputation. It uses a new material of that time called {\em polyurethane} to develop near-natural but artificial leg. Since the last fifty years, the Jaipur foot has been assisting thousands of physically challenged people to compete with ordinary people, bringing a smile on their faces. Durability, flexibility and convenience of use are some of the critical factors behind the successful and widespread use of Jaipur foot which was the result of innovative use of material technology.

The example of {\em Jaipur foot} illustrates two significant aspects of humanitarian engineering. First one is the use of technology, i.e. material technology in case of Jaipur foot and the second one is relieving pain or grief of underprivileged people.

The focus of this chapter is on blockchain technology, one of the emerging Information and Communication Technologies. The objective of this chapter is to illustrate how blockchain technology has promising applications to address the diverse needs of various developmental sectors as envisaged by the United Nations (UN) sustainable development program.

\begin{table}[t]
\begin{center}
\begin{tabular}{|p{0.5in}|p{1in}| p{4in}|}
\hline
Sr. No. & Development \newline Agencies & Activities \\ \hline
1 & UNDP & United Nations Development Program (UNDP) is one of the subunits of the United Nations headquartered in New York City. The primary role of the UNDP is to formulate policies, set priorities, and sponsor development programs to address the challenges faced by humanity. It does so by proactively working with all the member nations of United Nations Organizations (UNO)\cite{mccall2013undp}. \\ \hline

2 & UNWFP & United Nations World Food Program (UNWFP) is a branch of UNO raising the war against hunger and malnutrition. It provides food assistance to member countries so that the nutritional needs of the people leaving in poverty can be met. It aims to reduce child mortality, improve maternal health and to fight against diseases \cite{fao2015wfp}.\\ \hline
     
3 & WHO & World Health Organization (WHO) is a subsidiary of UNO dealing with matters of public health at the international level. It articulates policies for ensuring the highest possible health for individuals by identifying social and economic factors affecting health. It closely works with national government agencies and monitors the effective implementation of health policies \cite{world2002traditional}.\\ \hline
\end{tabular}
 \caption{\bf International Development Agencies and Their Role}\label{international}
\end{center}
\end{table}

\section{A Framework for understanding Humanitarian Engineering}

Before delving into details of technological alternatives for realizing humanitarian engineering, let us develop a framework to understand the dynamics of humanitarian engineering as a discipline. The three main pillars of humanitarian engineering are: (1) Stakeholders (2) Goals (3) Sectors for humanitarian engineering.
\begin{figure}[t]
\centering
    \includegraphics[width=4.5in]{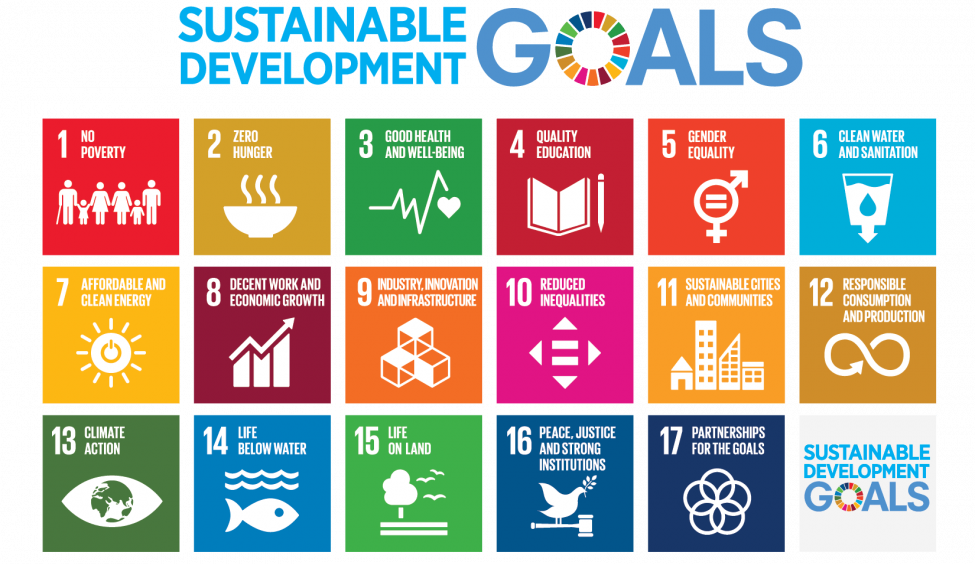}
    \caption{Sustainable Development Goals}\label{sdg} 
\end{figure}
\subsection{Stakeholders for Humanitarian Engineering}
The main stakeholders for humanitarian engineering projects are the people and organizations that play a critical role in the development of underprivileged communities. These are also known as {\em development agencies}. These agencies operate at different geographical levels from global to local level.

The first kinds of development agencies are {\em international development agencies}. Their primary function is to address the problems faced by humankind. These agencies set the agenda
for development programs, identify the problems, monitor and evaluate the performance of development programs, collect data and evidence to study the impact of development programs and to fund the developmental activities in various countries.

Table \ref{international} describes the role of three leading international development agencies. Besides this, the World Bank and Asian Development Banks (ADB) are some of the other international development agencies which sponsor developmental projects.

\begin{table}[t]
    \begin{center}
        \begin{tabular}{|p{1in}|p{2.5in}|p{2.5in}| } \hline
            Category & SDGs & Description \\ \hline 

            {\bf People} & (SDG 1) No Poverty,   (SDG 2) Zero Hunger,   (SDG 3) Good Health and Well-being,   (SDG 4) Quality Education,   (SDG 5) Gender Equality. & End poverty and hunger in all their forms and dimensions, ensure all human beings can fulfil their potential in dignity, equality and healthy environment. \\ \hline   

            {\bf Prosperity} & (SDG 8) Decent Work and Economic Growth. & Enjoy prosperous and fulfilling lives, economic, social, and technological harmonic progress. \\ \hline 

            {\bf Peace} & (SDG 10) Reducing Inequality,   (SDG 16) Peace, Justice, and Strong Institutions. & Foster peaceful, just and inclusive societies, free from fear and violence. \\ \hline 

            {\bf Partnership} & (SDG 17) Partnerships for the Goals. & Revitalized global Partnership, the participation of all countries, stakeholders and people. \\ \hline 

            {\bf Planet} & (SDG 6) Clean Water and Sanitation,   (SDG 7) Affordable and Clean Energy,   (SDG 9) Industry, Innovation, and Infrastructure,   (SDG 11) Sustainable Cities and Communities,  (SDG 12) Responsible Consumption and Production,   (SDG 13) Climate Action,   (SDG 14) Life Below Water,  (SDG 15) Life On Land. & Protect degradation through sustainable consumption, production, natural resource management, stakeholders and people. \\ \hline 
        \end{tabular}
        \caption{\bf Sustainable Development Goals}\label{sdg}
    \end{center}
\end{table}

Many countries in the world have development agencies at the national level to formulate country-specific development policies, align it with policies formulated by UNDP, and implement developmental projects. For example, in India, the National Initiative for Transforming India (NITI) Aayog, is one such organization.

Also, there are some regional development agencies whose primary role is to implement the developmental projects in coordination with national and international agencies. In a few countries, universities and academic institutes are slowly emerging as a regional knowledge centre to provide information necessary to get insights about the local developmental needs \cite{uni}.

\subsection{Sustainable Development Goals}
The Sustainable Development Goals (SDG) shown in Figure 1.1 capture the expectations and aspirations of people to live in a prosperous society without compromising on the needs of future generations. These SDGs are a set of ambitious goals put forward by the UN in the Year 2015 for the entire development of humanity \cite{assembly2015sustainable,nilsson2016policy}. These set of goals are linked to specific targets to be achieved by the end of the Year 2030. Table \ref{sdg} list out all seventeen goals categorized into five labels, namely People, Prosperity, Peace, Partnership and Planet \cite{wu2018information}.

Achieving these goals is an enormous task. Development agencies need to devise social, legal, financial and technological interventions. A development engineer responsible for technological interventions should know these SDGs so that project objectives and outcomes can be aligned with the SDGs. Knowledge of these SDGs is also essential for any engineer to become conscious of societal needs and environmental responsibilities.

\subsection{Sectors for Humanitarian Engineering}

The set of SDGs described in the last section aims for the overall development of humanity. The three main sectors which directly capture the needs of the human being responsible for achieving SDGs are Economic systems, Social systems and Environmental systems. However, some additional institutes and sectors are responsible for laying the operational and legal framework necessary for business transactions. A legal system which enacts the laws protecting the environment is one such example. This section reviews such business sectors responsible for achieving the target indicators set for each SDGs.

\begin{enumerate}
 \item {\bf Agriculture} The agriculture sector is the most crucial sector for sustainable development. The SDGs like {\em End Poverty} and {\em Zero Hunger} are directly related to agricultural production. Feeding the ever-growing population with sufficient food and nutrient is a huge task. As most of the rural population depends on income from agricultural produce, the goal of ending poverty is also correlated with agriculture. The sustainable agricultural production is constrained by factors like growing population, water shortages, declining soil fertility and climate change \cite{wu2018information}. Recently scientists have been exploring technologies such as Wireless sensor technologies\cite{culibrina2015smart} and ICT\cite{singh2015innovation} for Sustainable agricultural production

 \item {\bf Banking and Finance}
 Banking and financial markets drive economic development. It provides capital to start new businesses. It provides various avenues for income growth and wealth accumulation \cite{rioja2004finance}. The number of banks and non-banking financial institutes present in the community is one of the indicators of economic development. Trust in financial institutes, efficiency of processing business transactions (e.g., remittances, payment made to farmers on selling upon crops), diversity of financial instruments (e.g., crop insurance, loans, micro-credits) are some of the factors necessary for decent work and economic growth. Emerging technologies such as cryptocurrencies have to play a significant role in making business transactions simpler. 
 \item {\bf Education} Education brings changes in behaviour. Education makes people more knowledgeable and skilful to get decent jobs. Education brings awareness about the environment, and once social responsibility. It also has an indirect impact on the health of an individual, reduction in the rate of population growth. The task providing education to all is becoming more challengable because of a shortage of trained teachers, difficulties in providing better learning experiences to students on account of a shortage of playground in urban areas and oversized classrooms. Educators are gradually adopting technologies to overcome some of these challenges.
 \item {\bf Energy} The goal of {\em affordable and clean energy} (SDG 7), is directly related to the Energy sector. This sector also drives industrial development and economic growth. The demand for energy is increasing exponentially because of increasing population and wide-spread use of electro-mechanical devices to carry out routine works. Conventional energy generation methods that use natural resources such as coal and fossil fuels fail to meet this increasing energy demand. Hence use of non-conventional and renewable energy sources need to be increased on a large scale for sustainable development. In this context, it is required to provide technological solutions that would reduce energy demand, techniques for efficient energy production and replacement of conventional energy sources\cite{lund2007renewable}.
 \item {\bf E-Governance} E-governance is the use of Information and Communication Technologies (ICT) into day-to-day government processes. It is aimed to bring effectiveness in administration, transparency in the services provided by the government, and increasing participation of citizens in implementing government policies. The sector of E-governance is directly linked to Goal 16 and Goal 17. This sector plays a critical role in building accountable institutes necessary for enforcing peace and justice in societies (Goal 16). It also lays the technological infrastructure necessary to strengthen partnership at the global level required to implement various SDGs (Goal 17) \cite{ndou2004government,lim2016government}.
 
\begin{table}[t]
\begin{center}
    \begin{tabular}{|p{0.5in}|p{2in}|p{3.5in}|}
\hline
            {\bf Sr. No.} & {\bf Emerging Technologies} & {\bf Benefits of the technology} \\ \hline
            1 & Bioplastic, Bio-degradable plastic and Sustainable plastic & To Protect from climate change and prevent degradation of natural resources (SDG 13) \\ \hline

            2 & Renewable Energy Sources (e.g., Wind, Solar, and biofuels) & prevents degradation of natural resources, clean and green energy (SDG 7)\\ \hline

            3 & Electric Vehicle & Effective fuel consumption and pollution-free automobiles (SDG 7 and SDG 13)\\ \hline

            4 & Artificial Intelligence and Machine Learning & Personalized Medicine, outcomes-based public health and diagnosis of cancer, pneumonia and other diseases (SDG 3) \\ \hline

            5 & Communication Technologies (e.g., Mobile communication, Wireless communication, Smartphone, Internet) & Dissemination of knowledge between doctors, patients and caretakers (SDG 3) \\ \hline 

            6 & 3D Printing & Lowers cost of personalized medicine and organ transplant (SDG 3)\\ \hline

            7 & Genetic Technology & Facilitates precision medicine (SDG 3)\\ \hline

            8 & Digital Financial Technologies (e.g., Cryptocurrencies and Blockchain) & Supports micro-payments, create a trustworthy environment for business transactions (SDG 8 \& SDG 17)\\ \hline

\end{tabular}
\caption{\bf Emerging technologies for Sustainable Development} \label{tech}
\end{center}
\end{table}
 \item {\bf Environment Science and Engineering}
 The policymakers formulating the developmental goals have realized the drawbacks of uncontrolled development that took place in the last century. The economic growth that pollutes air and water that reduces the portion of forest and agricultural land, never satisfy the needs of the future generation. Hence the SDGs such as the provision of clean water and sanitation (SDG 6), actions for regulating climate change (SDG 13), and protecting life below water and on land (SDG 14 \& SDG 15) are directly linked to environmental science and engineering. Newer clean and green technologies need to be developed and adopted, especially in the areas of civil engineering, construction, water management, urban development, and for maintaining biodiversity \cite{clayton2018sustainability}.

 \item {\bf Health}
 To provide good health and well being are the sustainable goals which are directly linked to the health sector. Besides the lack of primary health care mechanisms, there exist many factors which affect the health and wellness of individuals. Some of these are the byproducts of uncontrolled development. For example, industrialization and modern urban centred life are causing stress and respiratory system-related diseases. Technologies such as information and communication, genetic engineering have been found useful in preventing, monitoring and diagnosis of diseases\cite{deen2015information}.
 
\end{enumerate}

\section{Technological Perspective of Humanitarian Engineering}
The sustainable development goals listed in the previous section are ambitious in terms of number, scope, and indicators used to measure the attainment of these goals. One of the effective ways to achieve the set targets is through innovative use of existing and emerging technologies. Table \ref{tech} lists some of the emerging technologies that will have a positive impact on sustainable development.

These technologies can help to combat climate change, degradation of natural resources, achieve good health and well being, to speed up economic growth and promote all-inclusive and equitable social development.

Data is one of the common threads across all the technologies listed in Table \ref{tech}. It has been envisaged that in the coming decade, data will drive sustainable development. At the same time, data-driven development is continuously raising many issues, such as security of information, privacy, and ethical concerns.

In the following sections, we discuss and evaluate one the emerging technology called blockchain technology in this context.

\section{A Blockchain Primer}

The necessity of blockchain technology can be understood through the knowledge of  potential and pitfalls of the Internet as a platform for business.

The Internet has introduced an information-centric model of business, and it has revolutionized the way people transact online. For example, the emergence of e-commerce sites (e.g., Amazon) has been attributed to the growth and widespread presence Internet.

The Internet has bridged the information gap that exists between a service provider and service consumer by creating a third-party for information exchange called intermediaries or agents or service providers. These agents which are e-commerce sites, hold the information about who sells what, i.e. seller's information and who wants what, i.e. buyers profile and their needs — thus bringing together consumers of services or goods with that of producers.

The advantages of doing online business are that it simplifies the process of business transactions, reduces the time required for businesses, and as a result, it has brought prosperity in the society. 

In the context of SDGs, the Internet as a platform for business has created various opportunities for decent work and economic growth. The Internet as a platform for communicating information has reduced the impact of natural disasters such as cyclones and spread of epidemics by timely disseminating useful information. The Internet as a learning platform has increased the accessibility of education fostering the goal of education to all. 

Despite the various benefits of the Internet, it has always remained an unreliable platform to share valuable personal information because of its mediator-centric or client-server model for information exchange. As shown in Figure \ref{fig:sub1}, personal information is exchanged through mediators or servers. A server or mediator may be a payment gateway or an e-commerce site. The information shared with such sites is always susceptible to breach of security and privacy attacks.

\begin{figure}
\centering

\begin{tabular}{cc}
  \includegraphics[scale=0.5]{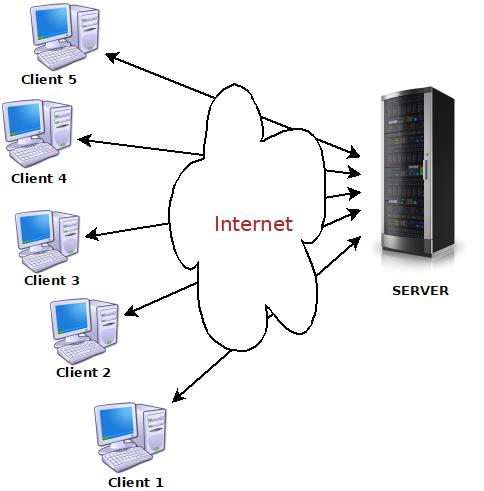} &
  \includegraphics[scale=0.8]{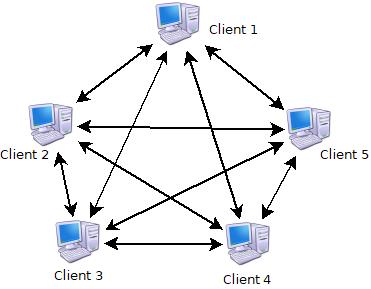} \\ 
 (a) Client Server System & (b) Decentralized Systems
\end{tabular}

 \caption{Client-Server versus Decentralized Systems}
\label{fig:test}
\end{figure}

The emerging blockchain technology removes these pitfalls by laying a trust layer on top of the existing Internet technology. It replaces the mediator-centric model of information exchange with the peer-to-peer model or decentralized model, as shown in Figure \ref{fig:sub2} of information exchange. It transforms the Internet into a trustworthy platform for doing business when transacting parties do not trust each other. It eliminates the role of mediator responsible for authenticating the identities of transacting parties. Table \ref{tab1} differentiates the client/server and decentralized model of interactions.

\begin{table}[t]
\begin{center}
\begin{tabular}{p{1in}p{2in}p{2in}}
    \\ \hline
    ~ & Client/Server Architecture & Decentralized Architecture \\ \hline \\
    Mode of Communication & Client and  Server communicate through each other via Inter-networking protocols & {\em Peer-to-Peer} communication via inter-networking protocols. \\
    Components & Nodes, Server, Communication Links & Nodes, Communication Links \\
    Architecture Style & Layered & Object-based \\
    Availability & Need to achieve high availability & No need to achieve high availability \\
    Servers & Only single server is present to serve clients & Servers are distributed in a system. \\
    Privacy & It is easy to determine who is handling the content & Difficult to determine who is handling the content \\
    Failure & Bottleneck as the sever fails down & No bottleneck as services can be provided through other nodes in case of failure. \\
    Cost & Implementation cost is high & Implementation cost is negligible \\ \hline
    \\
    
\end{tabular}
\caption{Difference Between Client/Server and Decentralized Architecture}\label{tab1}
\end{center} 
\end{table}
Initially emerged as a platform to exchange digital currency over the Internet, now the blockchain technology gradually emerging as a general-purpose platform for doing business over the Internet. Due to potential applications of blockchain technology in various Fields, UN has included it as one of the frontier technologies to realize SDGs \cite{ftech}. 

This section provides an overview of essential elements of blockchain and how it achieves the various quality attributes that make it as one of the promising technology. The blockchain technology can be understood at the conceptual level and specific instance level. The Bitcoin, Ethereum, and Hyperledger are a few common examples of specific blockchain. This section reviews the blockchain technology at the conceptual level. The four fundamental concepts common across the blockchain implementation are \cite{dinh2018}:
(i) Distributed Ledger,
(ii) Cryptography,
(iii) Consensus Protocols, and
(iv) Smart Contracts

\subsection{Distributed Ledger} 

In a conventional sense, ledgers are the registers or logbooks employed for account-keeping or book-keeping operations. Similarly, in the context of a blockchain-based information system, ledgers are the databases storing up-to-date information about business transactions. These are distributed among all the nodes participating in the network. In a blockchain environment, ledgers are not stored at a central place. They are distributed among all the nodes.  So multiple copies of a ledger exist in a business network. Hence these are referred to as distributed ledgers. When a node in a network updates its local copy, all other nodes synchronize their copy with the updated one.  Hence, each copy is consistent with each other.

These ledgers are used to store information about valuable assets. In the Bitcoin implementation, the first blockchain-based system, ledgers are used to store digital currencies. It may be used to store information about other valuable assets such as land records, diamonds, student's academic credentials and others.

\begin{figure}[h]
\centering
    \includegraphics[width=4.5in]{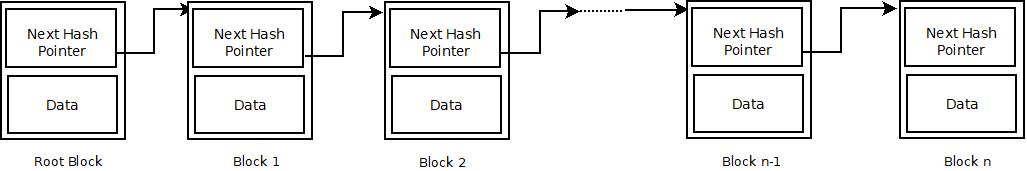}
    \caption{Blockchain}
    \label{bc}
\end{figure}

In a blockchain-based information system, records in a distributed ledgers are arranged in a chain-format, as shown in Figure \ref{bc} for storage purpose. Here, multiple transactions related to an asset are grouped in a block. The $(n+1)^{th}$ block in the chain links to the $n^{th}$ block and the $n^{th}$ block links to the $(n-1)^{th}$ block and so on. The first block in a chain is called a genesis block or the root block. Due to this peculiar storage arrangement, the distributed ledgers are also known as Blockchain. Here it is worth to note that all blockchain-based systems contain distributed ledgers but not all distributed ledgers employ blockchain-based storage mechanism. The blockchain data structure permits only append of new records. Updating and deletion of records are not permissible.

The most critical design feature of blockchain-based information system is the use of hash pointers instead of physical memory based pointers to link blocks in a chain. A hash pointer is a message digest calculated from the information content of a block. Whenever a node attempts to tamper the information content, a small change in the information leads to a ripple effect of changes in hash-pointers — making it impossible to change the information once it has been recorded in the blockchain. 

Facilitating mediator-less business transactions and supporting immutability of stored information are the two significant quality attributes associated with blockchain-based information systems. These quality attributes are derived from replicating ledgers on all the nodes in a network and linking blocks in a chain through hash pointers.

Typically, blockchains are of two kinds based on how blockchains are accessed, i.e. private and public blockchains. In a public blockchain, any node can join and leave the network and validate the business transactions. While in the case of private blockchain, the network is small and requires permission to join and leave the network. Hence private blockchains are also known as permissioned blockchain, and public blockchain is known as permission-less blockchain. For example, Bitcoin is a public blockchain, and Hyperledger is an example of private blockchain.

\subsection{Cryptography}
Blockchain technology makes heavy use of cryptographic functions to assure trust among the users transacting over a blockchain-based business network. A typical business network includes many un-trustworthy elements. In a conventional banking domain, an agent issuing the check without having sufficient balance in the account, or an agent forging a signature are typical examples. In a digitized economy, these challenges are aggravated because of information transfer over an unreliable communication medium. Hence, cryptographic functions, a set of mathematical functions, are used to encode messages to assure the security of information in a network containing malicious or untrustworthy agents. These cryptographic functions address various purposes. Some of them are:

1) {\em Authenticating the identity of agents involved in a business transaction}:\newline Blockchain-based systems use a kind of asymmetric key cryptography. These protocols use two different keys called public and private keys. The public keys are open and used as addresses for performing business transactions while private keys are secret and used for validating the transactions. SHA-256 (e.g., Bitcoin) and ECDSA (e.g., Hyperledger) are some of the cryptographic protocols used for this purpose. The private blockchain use another service called membership service, which authenticates the identity for business transactions.

2) {\em Ensuring Privacy}: Maintaining the privacy of transactions is a challenge, especially in public blockchains (e.g., Bitcoin). In such systems, transactions are possible to trace to real-life identities. Advanced cryptography-based techniques such as cryptographic mixers (e.g., Zerocoin) and Zero-Knowledge proof (e.g., Zerocash) have been found useful to address this challenge.

Cryptographic functions such as digital signature are also used to authenticate a particular transaction.

\subsection{Consensus Protocols}
In decentralized systems, agreeing upon the global state of the transaction is a challenge. In a centralized system, this is not an issue because only one copy of transaction history is present at the central authority (e.g., Banks main Server machine). Blockchain being a decentralized system, holds multiple replicas of transactions at several nodes. Agreeing upon the unique state of the transaction is an issue which is solved by executing a consensus process involving all the nodes in the system. This process is typically carried out in three stages. In the first phase, a node is elected/selected as a leader node to decide upon a unique state. In the second stage, transactions are validated. In the third stage, transactions are committed. A variety of consensus algorithms exists in blockchain-based system. These are often compared based upon how scalable the algorithm is and several malicious nodes it tolerates. The Proof-of-Work (PoW) algorithm used in Bitcoin is one example of the consensus protocol. It selects the leader node responsible for deciding upon a global state by solving a cryptographic puzzle. It takes about 10 minutes for solving the puzzle requiring extensive computational work and much electric energy. It can work in the presence of 50\% of malicious nodes in the network.
The Proof-of-Stake (PoS) is another consensus protocol in which a leader is selected with the highest stakes in the network. It has been found as scalable as compared to PoW, and it also works in the presence of 50\% of malicious nodes in the network.

The Practical Byzantine Fault Tolerant (PBFT) is the third example of consensus protocol which has been found scalable and works in the presence of 66\% (2/3) malicious nodes in the network.
\vspace{-0.25in}
\subsection{Smart-Contracts}
Smart-contracts are the most significant element in the blockchain-based system because it provides configuring the behaviour of such systems. Blockchain programmers can customize the working of blockchain systems by writing programs called {\em Smart-Contract}. The smart contracts are scripts which are executed when a specific event occurs in a system. For example, in the context of Bitcoin, a coin may be released when more than one signatures are validated, or when miners solve a cryptographic puzzle.

These scripts can be written in a native language provided by blockchain systems or general-purpose programmable language. For example, Bitcoin provides a simple and less expressive native language to write a smart contract while Ethereum provides a Turing complete native language called Solidity to write smart contracts. In Hyperledger, blockchain programmers can write a smart contract in a general-purpose language such as Java/Go.
 
\section{Sectoral Applications of Blockchain Technology}

\subsection{Blockchain Technology in Agriculture: \label{agri}}
Numerous business factors motivate the adoption of blockchain technology in Agriculture sector. 

First and the topmost key-factor behind the adoption of  blockchain technology in agriculture is to eliminate bad actors and poor processes involved in the food supply chain. For example, {\em AgriDigital} ($agridigital.io$) is a cloud-based platform that is designed using blockchain technology to provide an efficient interface for the agriculture supply chain.

The second key factor is to create trust among consumers and retailers about the originality and authenticity of food products. For example, $ripe.io$ is an organization which provides a blockchain-based platform with this sole purpose.

Blockchain technology has the potential to open up new market areas for food producers, specifically farmers from the developing nations. For example, {\em AgriLedger} ($agriledger.io$) adopts the distributed ledger technology to open up new market areas for the farmers.

We broadly classify the typical use-cases of blockchain technology in the agriculture sector as below:
\begin{enumerate}
    \item {\bf Crowd-funding for the development of agricultural products} Entrepreneurs and start-up organizations have been increasingly using blockchain as crypto-currencies or digitized tokens to raise capital for their entrepreneurial ventures. This mode of application is generally known as Initial Coin Offerings (ICO) \cite{catalini2018initial}. 
    
    The ICOs are a type of digitized financial instruments which can be monetized through crypto-currencies or fiat currencies. The $NagriCoin$ ($nagricoin.io$) is one such application, which has raised capital to develop smart fertilizers that stimulate plant growth. 
    
    \item {\bf Development of agricultural supply chain management} Various business factors drive the use of blockchain technology for the agricultural supply chain management. First, paying the farmers their due share in food production is the key motivating factor behind the adoption of blockchain technology for the food supply chain. The second most important factor is to address the concerns about food safety in the minds of consumers. 
    For example, the $AgriDigital$ is a cloud-based supply chain management platform primarily supporting the supply chain for grains. 
    
    The objective of such platforms is to effectively connect the various operators in the supply chain such as farmers, distributors, storage operators, retailers and consumers. 
    
    Creation of digital assets representing physical products, recording and tracking their flow within the supply chain are the main functionalities offered by such platforms. They implement it by providing a set of smart-contracts executing over a blockchain implementation such as Ethereum.
    
    \item {\bf Tracking and Traceablity}
     Assuring about the quality of food products and agricultural commodities is one of the main challenges faced by agriculture and food processing industries \cite{kamilaris2019rise}. Consumers expect authenticity about food products in terms of composition, origin, and purity.
    These expectations are valid when the food-products are costly, consumed at a distant place far from its origin, and they are directly linked to the health and safety of consumers.
    
    Providing accurate and verifiable information about the composition and origin of the food product is one of the solutions to address this challenge. The blockchain technology plays this role precisely while tracking and tracing the origin of food production.
    
    The blockchain-based solutions to address this problem are typically built around the immutability feature of the blockchain along with other technologies such as DNA marking, RFID tagging, QR coding, and isotope testing.
    
    \item {\bf Blockchain-based financial products for farmers}
    Conventional financial institutes such as banks, insurance companies and cooperative credit societies face the challenge of including small-farmers in the mainstream financial sectors. This situation leads to undue exploitation of farmers by illegal money lenders and traders.
    
    Blockchain, as a technology initially emerged as a payment system through digital currencies (e.g., Bitcoin) can be used to provide financial products specially designed for farmers.
    
   The $AgriWallet$ is one such example of saving scheme designed and implemented by COIN22 ($coin22.com$) for smallholder farmers to manage the risk associated with droughts, floods and low-yield of crops \cite{fao}.
   
   Also, the use of blockchain technology is currently being explored by many companies and government agencies to provide agriculture insurance, micro-credit and subsidy transfer \cite{bermeo2018blockchain}.
\end {enumerate}
Mapping of the physical commodity to the digital artefact is one of the challenges faced while designing such platforms. This challenge can be addressed by using the Internet-of-Things (IoT) and various sensors along with blockchain technology. The sustainable goal of ensuring zero hunger (SDG 2) and no poverty (SDG 1) is indirectly supported by extensive adoption blockchain technology in the agriculture sector.

\subsection{Blockchain for Financial Services:}

The factors such as to eliminate the role of mediators between two transacting parties, to reduce transaction time, and to simplify the transaction process motivates the use of blockchain in providing financial services.

In such scenarios, blockchain technology mainly plays the role of validating transactions. 
It helps to reduce the transaction charges by offering an efficient and secure way to transfer the funds either within a country or between two different countries. 

The use-cases of blockchain technology in finical sectors belong mainly to two categories, first, in the area of payment and transfer, and other record-keeping and verification.

For example, the \textit{Mojaloop} ($mojaloop.io$) is one of the examples of the block-chain-based payment transfer platform developed for Gates Foundation\newline ($gatesfoundation.org$) to connect financial service-providers and customers from all over the world. In addition to offering efficient payment mechanism, blockchain technology reduces the possibilities of frauds such as corruption during the disbursement of funds.  

Secondly, the features of blockchain technology, such as distributed ledger and smart contracts, have been in use for effective aid distribution. Each step involved during the aid distribution from donation to receipt of the amount can be tracked and monitored using distributed ledgers. Donations are released upon meeting specific criteria and occurrence of pre-specified events encoded in the smart contracts. For example, $Amply$ \cite{bi} is a blockchain-based protocol developed by IXO Foundation which links the release of government subsidies to the attendance of children in a school.

Applications of blockchain technology in the banking and finance sectors enable the attainment of sustainable development goals such as SDG 1, SDG 9, \& SDG 8. For example, the World Food Program (WFP) used the blockchain technology to provide food assistance to refugees from Syria.  
        
\subsection{Blockchain in Education:}

The information and communication technologies have revolutionized the mode of delivering education. It has realized the goal of open education to all and created multiple ways to equip oneself and upgrade the skill set. The blockchain technology is now providing value-added services in the education sector, which includes credit transfer, issuing a transcript, digital certificates, and payment for online courses. 

We use the following categories to classify the uses of blockchain technology in the education sector.
\begin{enumerate}
    \item {\bf Record Verification and Validation} In this mode, developers have been using blockchain as distributed ledgers which store and verify academic credentials of students. For example, $BlockCert$ ($https://www.blockcerts.org/$) is a set of tools that provides the functionality to create, store and validate blockchain-based digital certificates. 
    \item {\bf Information Sharing} Universities need to share and transfer the academic credentials of a student when a student migrates from one university to another. In this mode of application, blockchain technology enables effective transfer of academic information between two universities.  For example,  in \cite{turkanovic2018eductx} describes the design of one such platform called $EduCTX$ that enables credit transfer among institutes.
    \item {\bf Identity Management} Online learners use multiple applications and services such as Coursera, Google Classroom, and note-taking apps (e.g., Keep). Securely sharing and authenticating identity credentials across the applications remain a big challenge. Blockchain technology addresses this challenge by providing a unique identifier accessible and verifiable across the applications. For example, $BlockStack$ ($https://blockstack.org/$) one such cloud-based identity management service.

These applications of blockchain technology have realized the goal of providing equal learning opportunities to everyone i.e. (SDG 4).
\end{enumerate}

\subsection{Blockchain in Energy Sector:}
The energy sector worldwide is undergoing structural as well as functional changes. The decentralized network of renewable energy generators have been gradually expanding, which is currently dominated by the conventional centralized grid structure. The functionalities of the conventional power grid are also not only limited to the functions of power generation and distribution but providing additional services such as analyzing the data generated through the deployment of smart devices and offering intelligent services to consumers.

The blockchain technology is the driving force behind all these changes. The features of blockchain, namely distributed ledger and smart contracts, are being increasingly used to provide additional services. Andoni M, Robu V, Flynn D, et al. present a comprehensive survey of such services in \cite{andoni2019blockchain}. Some of these services are:

\begin{enumerate}
    \item {\bf Wholesale Energy Trading and Supply} The technologies such as distributed ledgers and smart contract are used to simplify the business processes involved in energy trading and supply. In this context, developers use blockchain technology primarily for record-keeping, autonomous agents executing business logic, and for payment and transfers. Elimination of intermediaries and reduced transaction times are the benefits gained by adopting the blockchain-based approach.
    \item {\bf Imbalance Settlement} It is a process executed by grid operators. It aims to remove discrepancies between energy units sold by energy suppliers and the energy sold by distributors to the consumers. Blockchain technology provides an efficient solution to this problem which normally takes months for settlement. Here also, distributed ledgers are used to keep track of energy generation and consumption.
    \item{\bf P2P Trading} Peer-to-Peer (P2P) trading is a form of trading in which energy providers or generators directly sell the energy units to consumers. Thus eliminating the role of energy distributors. Here blockchain technology is used as a platform that facilitates energy trading in-exchange of digital currencies. 
\end{enumerate}

Providing affordable access to reliable and modern energy (SDG 7) is one of the sustainable goals that blockchain technology realizes through the use-cases, as mentioned above. Further blockchain technology is providing an effective platform for energy trading in micro-grids and community-owned energy systems.

\subsection{Blockchain for E-Governance:}

Government agencies face the challenge of providing transparent and citizen-centric services. 
These agencies are looking at blockchain technology as a medium to improve the quality of their service.
These agencies have been adopting blockchain for multiple application areas. Some of which are listed below:
\begin{enumerate}
    \item {\bf A Platform for offering integrated services} The government of Estonia has been using blockchain as a platform to provide more than a thousand services to its citizen. Here, the distributed ledgers are used for logging and auditing each transaction of citizens with the government. The government of Estonia has issued digital-ID and signatures to all citizens for this purpose. The use of blockchain has brought transparency and trust in the services offered by the Estonian government.
    \item{\bf E-voting} Democratic societies face the challenge of conducting free and fair elections. Conventional ballot-paper based elections are prone to be rigged and manipulated. Blockchain technology is currently being explored as an alternative to address this challenge. The $Votem$ is a cloud service that uses blockchain to securely cast votes by mobile or remote users. The system uses distributed ledger technology to record and verify the votes. 
    \item {\bf Land Records} Recording the information of landholding and ownership is one of the obvious usages of blockchain technology. The recording of such information in the immutable registry and for archival purpose is essential to avoid land-related frauds. Some countries, such as India and Sweden, have initiated the use of blockchain technology for land records. In cities such as Chicago, civic administrators have been using blockchain technology to automate the complete workflow involved in land registry.
\end{enumerate}

The blockchain technology, in some cases, have been directly realizing the goal of Peace, Justice, and Strong Institutions (SDG 16). While in other cases, it has been indirectly supporting the goal of equitable development (SDG 5).

\subsection{Applications of Blockchain in Environment and Climate Change:}
Developers have started thinking about innovative applications of blockchain technology in the areas of environmental protection and climate change.  Some of these novel applications include:
\begin{enumerate}
    \item {\bf Incentive Mechanism for Recycling Project} The blockchain technology is useful in rewarding digitized tokens or crypto-currencies for joining in the recycling project. A digital token can be exchanged for the deposit of plastic wastage.
     \item {\bf Enforcing Environmental Treaties} Government agencies face the challenge of enforcing environmental obligations while sanctioning developmental project. Developers have started applying the feature of the smart contract embedded in various blockchain platforms for this purpose. 
     \item{\bf To implement Carbon Tax} The carbon footprint is a measure used for assessing carbon-dioxide emission caused by an individual, event or an organization. The blockchain technology is useful to track and record the carbon footprint. Further, it can levy a tax based on carbon footprint or build a reputation system to provide incentives for low carbon footprint products and services.
    \end{enumerate}

These applications illustrate the potential of blockchain technology in protecting the environment and ensure responsible consumption and production (SDG 12 and SDG 13).

\subsection{Blockchain in Health Sector:}
The applications of blockchain technology in health sector belong to the following categories.
\begin{enumerate}
    \item {\bf Record keeping and ensuring confidentiality of information} The use of\newline blockchain technology in the health sector is driven by the motive to overcome the negligence of health-related documents and privacy of these documents (e.g., medical prescriptions, and health reports). Preserving this information is not only useful to an individual but in carrying out research related to genetic diseases.
    All such documents are valuable to medical researchers, health care providers, public health authorities and health insurance providers. 
    
    Blockchain technology provides secure access to store existing medical reports. These reports are made available on-demand through various blockchain platforms. One of the typical applications is MedRec\cite{ekblaw2016case}, which stores Electronic Medical Records (EMRs) on blockchain. This information is accessible in a secure way to various stakeholders, such as medical researchers and public health authorities. 
    
    \item{\bf Drug Supply Chain and Traceability}
    In the supply chain of drugs and medicinal products, it needs to track the initial details of medical products while it is in transit. The \textit{Modum.io}, a European company, has implemented a hardware sensor-based technology to monitor the temperature of drugs in pharmaceutical supply-chain. In such applications, drug-related information exchanged between drug manufacturers, wholesalers, pharmacists, and the patients are recorded and tracked to authenticate the originality of the drug. 
    The features like a distributed ledger and smart contracts are used to implement these functionalities.
\end{enumerate}

The combination of technologies such as Artificial Intelligence, Blockchain and Internet-of-Things are providing various solutions and products to realize the goal of smart technologies for a healthy life (SDG 3).

\section{A Cost-Benefit Analysis of Adopting Technology}
The blockchain technology is an emerging technology. It is yet to reach the level of maturity of many internet-based technologies such as web applications and mobile applications. In this scenario, it becomes essential to know the costs and benefits associated with the deployment of blockchain-based solutions.  From the humanitarian engineering point of view, it becomes more vital because it offers numerous opportunities to realize SDGs and will have an impact on all aspect of development.

The strengths of blockchain technologies include:
\begin{enumerate}
    \item {\bf It provides a platform to build trustworthy services} Many government and development agencies face the challenge of creating trust offered by their services which may include aid transfer. Blockchain technology addresses this challenge and enables the strengthening of institutes and services.
    \item {\bf It provides a platform for open market} As observed in Section \ref{agri},  blockchain technology is creating an open platform to sell and buy products without any mediators. Thus it has the potential to bring small producers in the mainstream global market leading to multiple avenues for prosperity and growth.
    \item{\bf It simplifies business processes and transactions} The feature of {\em smart-contract} in blockchain technology allows to code the business logic and execute it upon the occurrence of a specific event. For example, the release of payment upon successful delivery of goods. Smart-contracts are an effective mechanism to implement business logic when business processes are well defined, and it involves multiple business operators crossing the organizational boundaries.
\end{enumerate}
The following risks associated with the blockchain-based services need to be considered while implementing services on top of it.
\begin{enumerate}
    \item {\bf No Central Authority means greater responsibility on users} A blockchain system, by its design, is a decentralized system. It means there is no central authority to contact in case of loss of public keys or loss of passwords. Hence restoring users credentials becomes impossible in such events. Also, the data entered in a system is immutable one and needs to be entered carefully.  The implications of all these architectural elements put a greater responsibility on users to be careful.
\item {\bf High Energy Requirement of some Consensus Protocols}
blockchain-based systems use a consensus protocol to validate a business transaction. Some consensus protocols such as Proof-of-Work (PoW) used in Bitcoin have been found ineffective in terms of energy consumption. It needs a high amount of electric power to validate a business transaction. Hence this cost needs to be considered when one selects a public blockchain for implementation.
\item {\bf Maturity of code to replace business logic} Blockchain-based systems use smart contracts to encode business logic and well-defined business rules. These contracts are automatically executed, and it replaces the need for human interventions. However, the smart contracts thus implemented need to be tested rigorously and formally verified. Otherwise, it will lead to a situation called {\em hard-fork} of blockchain i.e. replacement of older code with newer code for the smart contract, which is a costly affair.
\item{\bf Legislation related to data privacy} In most of the countries, the use of public blockchain such as Bitcoin is legally prohibited for payments systems because of lack of provision to trace the identity of transacting parties. Also, blockchain-based systems use immutable distributed ledgers which means they are {\em write-once- never-forget} systems. Such systems violate the privacy principle of right-to-forget. Such legal and ethical concerns need to be considered before adopting a blockchain-based approach.
  
\end{enumerate}

\section{Conclusion}
The Chapter reviews the existing and potential applications of blockchain technology for humanitarian engineering. The scope of humanitarian engineering is vast. It includes every aspect of humanity, from agriculture to finance. Information and Communication Technologies (ICT) have played a vital role to address the challenges in these sectors by sharing and exchanging information. However, many of the challenges need to be addressed, such as information privacy and security, which require more effective solutions. Humanitarian engineers are looking towards blockchain technology as a potential solution to address many of these challenges.

The Chapter defines the scope of humanitarian engineering actions in terms of achieving the Sustainable Development Goals (SDG) formulated by UNO. It identifies seven sectors crucial for the attainment of SDGs. 
The Chapter further describes the challenges in these sectors, which can be addressed by the use of blockchain technology. 

In general, across all these sectors, blockchain technology provides an effective mechanism to track and record information exchanged in a chain of business transacting entities. It automates the business workflow by reducing human intervention. It opens up new global market opportunities by creating a sound digital payment system crossing the boundaries of nations.

Blockchain being a technology yet to achieve the maturity level of web and mobile technologies, the deployment of blockchain-based solutions come with risks and adverse consequences. The Chapter briefly illustrates such implementation risks by doing cost-benefit analysis from a humanitarian engineering point of view.

Two main contributions of the Chapter include: (1) Identifying the scope of human humanitarian engineering in terms of SDGs and sectors responsible for the achievement of these goals. (2) A functional evaluation of blockchain technology by describing potential and existing applications of blockchain for achieving SDGs.

 The functional evaluation presented in the Chapter highlights the promising role for blockchain technology to build a global society free from hunger, deceases, and with equal opportunities for growth and decent work. However, for more  theoretical  and foundational role of technologies and Engineers in a society one can  refer \cite{ESJ}.

\bibliographystyle{vancouver-modified}
\bibliography{chapter}

\end{document}